\documentclass[10pt, conference]{IEEEtran}

\makeatletter
\def\normalsize{\@setfontsize{\normalsize}{9.7bp}{12.00pt}}
\normalsize
\makeatother

\usepackage{psfrag, epsfig}
\usepackage{array}
\usepackage{epsfig}
\usepackage{multicol}
\usepackage{multirow}
\usepackage{amsmath}
\usepackage{amssymb}
\usepackage{threeparttable}
\usepackage{bigstrut}
\usepackage{amssymb}
\usepackage[nolist , nohyperlinks]{acronym}
\usepackage{graphicx}
\usepackage{subfigure}
\usepackage{makecell}
\usepackage{siunitx}
\usepackage{hyperref}

\usepackage{algpseudocode}
\usepackage{algorithm}

\newcommand{\ie}{i.e.}
\newcommand{\eg}{e.g.}
\newcommand{\fig}{Fig.\,}
\newcommand{\scn}{Sec.\,}
\newcommand{\tbl}{Table\,}

\graphicspath{{../figure/}}

\algnewcommand{\LineComment}[1]{\State \(\triangleright\) #1}

\usepackage{color}

\begin{document}

\title{Millimeter-Wave Massive MIMO Testbed with Hybrid Beamforming}
\author{
    \IEEEauthorblockN{MinKeun Chung$^{1}$, Liang Liu$^{1}$, Andreas Johansson$^{1}$, Martin Nilsson$^{1}$,\\
        Olof Zander$^{2}$, Zhinong Ying$^{2}$, 
        Fredrik Tufvesson$^{1}$, and Ove Edfors$^{1}$\\}
    \IEEEauthorblockA{$^{1}$Department of Electrical and Information Technology, Lund University, Sweden\\
    $^{2}$Sony Research Center Lund, Sweden\\
    firstname.lastname@\{eit.lth.se, sony.com\}}\\
    }

\maketitle

\begin{abstract}
Massive multiple-input multiple-out~(MIMO) technology is vital in millimeter-wave~(mmWave) bands to obtain large array gains. However, there are practical challenges, such as high hardware cost and power consumption in such systems. A promising solution to these problems is to adopt a hybrid beamforming architecture. This architecture has a much lower number of transceiver~(TRx) chains than the total antenna number, resulting in cost- and energy-efficient systems. In this paper, we present a real-time mmWave~(\SI{28}{\GHz}) massive MIMO testbed with hybrid beamforming. This testbed has a 64-antenna/16-TRx unit for beam-selection, which can be expanded to larger array sizes in a modular way. For testing everything from baseband processing algorithms to scheduling and beam-selection in real propagation environments, we extend the capability of an existing 100-antenna/100-TRx massive MIMO testbed~(below \SI{6}{\GHz}), built upon software-defined radio technology, to a flexible mmWave massive MIMO system. 
\end{abstract}

\begin{IEEEkeywords}
Beam-selection, beamforming, massive multiple-input multiple-out~(MIMO), millimeter-wave~(mmWave), testbed.
\end{IEEEkeywords}

\acresetall
\section{Introduction}
Massive \ac{MIMO} is a promising \ac{MU}-\ac{MIMO} technology where each \ac{BS} is equipped with an excess number of antennas, compared to the number of \acp{UE}, \eg, a few hundred \ac{BS} antennas simultaneously serving tens of \acp{UE}. The concept of massive \ac{MIMO} has been demonstrated to achieve an order-of-magnitude higher spectral efficiency with practical acquisition of \ac{CSI}, as compared to conventional small-scale \ac{MIMO} technology~\cite{mar10,rusek13}. In recent years, the development of massive \ac{MIMO} prototype systems, operating below-\SI{6}{\GHz}, has been carried out for proof-of-concept and performance evaluation under real-world conditions~\cite{shep12,mal17}.

Another key approach to enhance the network capacity is the operation in \ac{mmWave} bands, \ie, \SI{30}{\GHz}--\SI{300}{\GHz}~\cite{rap13}. It provides an order-or-magnitude more spectrum than we ever had access to. At the \ac{mmWave} bands, a large-scale antenna system, \ie, massive \ac{MIMO}, is imperative to obtain sufficient \ac{SNR} due to its high \ac{FSPL}~\cite{roh14}. However, there are fundamental differences between the design and implementation of massive \ac{MIMO} below-\SI{6}{\GHz} and at \ac{mmWave} frequencies. The main differences are summarized as follows:
\begin{itemize}
\item The architectures: the small wavelength at \ac{mmWave} frequencies enables a large number of antennas in a small physical size. However, the current high cost and power consumption of the \ac{TRx} chains at \ac{mmWave} frequencies make a fully-digital processing approach prohibitive. Hybrid analog and digital beamforming can be an alternative architecture for \ac{mmWave} massive \ac{MIMO} systems~\cite{han15, minkeun20_4}. This architecture has a much lower number of \ac{TRx} chains than the total number of antennas.
\item The propagation channels: propagation environments have a different effect on smaller wavelength signals. For example, diffraction, scattering, and penetration losses. It leads into different statistics of both small-scale and large-scale variations~\cite{shafi18}.  
\item The baseband processing algorithms: depend on hardware, as well as channel characteristics. As compared with below-\SI{6}{\GHz} systems,  the \ac{mmWave} system is more sensitive to hardware impairments, such as phase noise, \ac{PA} nonlinearities~\cite{minkeun20_7, minkeun20_10}. Thus, baseband processing algorithms for impairment estimation and compensation are crucial in \ac{mmWave} systems.
\end{itemize}

\begin{figure}[t!]
\centering
\includegraphics[width = 3.54in]{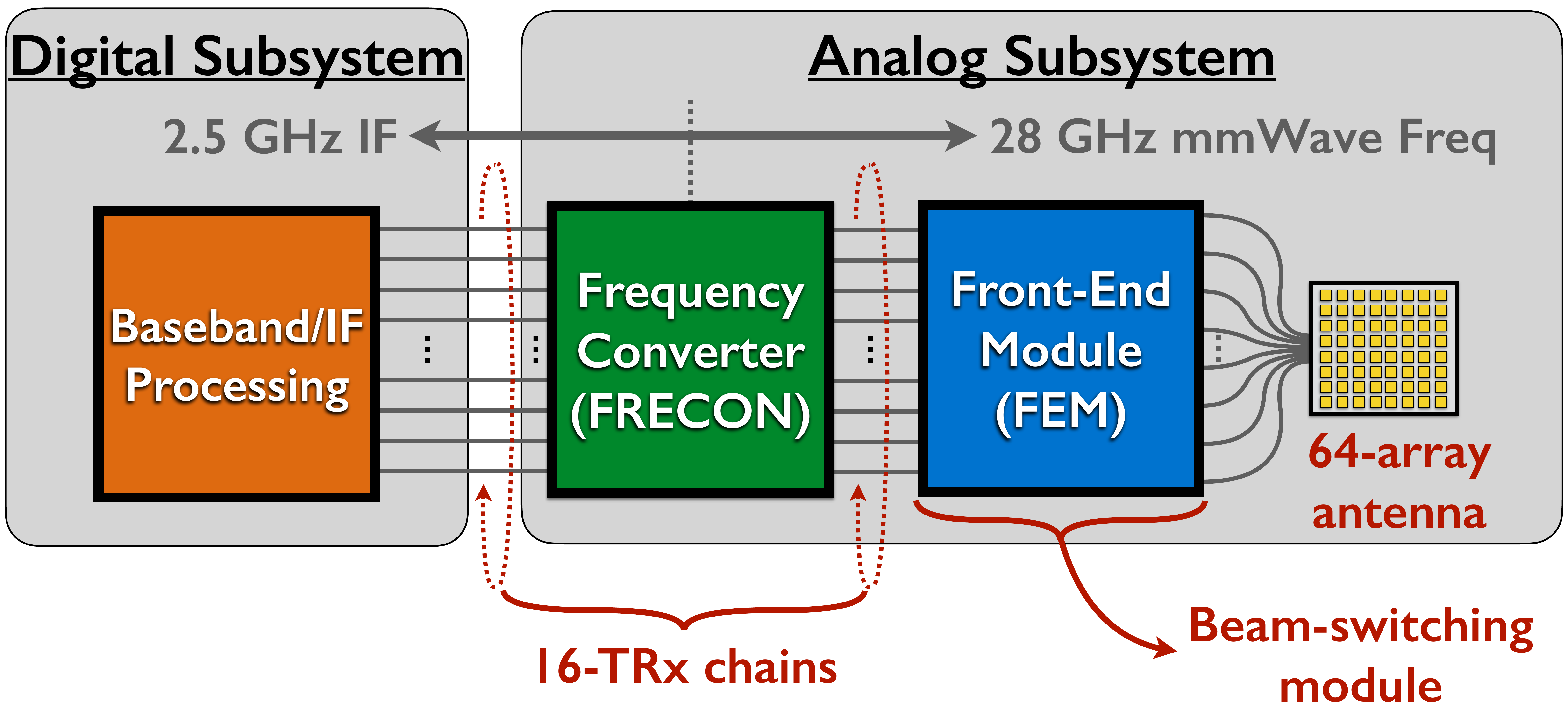}
\caption{System overview of our proposed mmWave massive MIMO testbed.}
\label{fig:Sys_testbed}
\vspace*{-0.25 cm}
\end{figure} 

Based on these differences, for testing everything from baseband processing algorithms to scheduling in new environments, we extend the capability of an existing 100-antenna/100-\ac{TRx} massive \ac{MIMO} testbed (below \SI{6}{\GHz}), built upon \ac{SDR} technology, to a flexible \ac{mmWave} massive \ac{MIMO} system. Recently, we have demonstrated this real-time \ac{mmWave} massive MIMO system at IEEE Wireless Communication and Networking Conference in 2020~\cite{minkeun20_4}.

\begin{figure*}[t!]
\centering
\includegraphics[width = 6.35in]{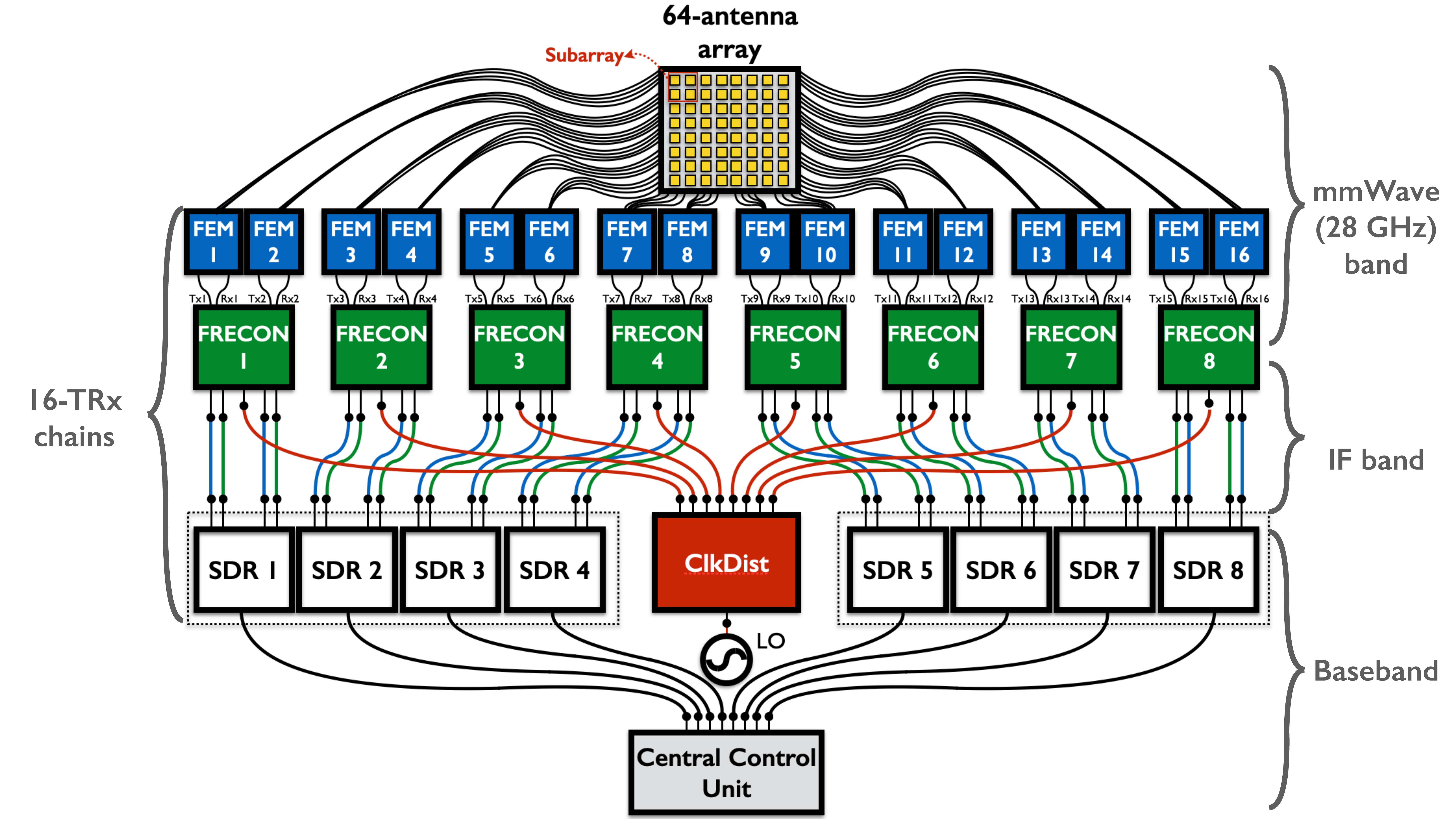}
\caption{An architecture of 64-antenna/16-TRx hybrid beamforming testbed. DC control signals from each SDR, \ie, for TDD and beam switching, are delivered to FRECON~(only TDD switching signal) and FEM (both). For simplicity, the routes for the DC control signals are omitted in this figure.}
\label{fig:Arch_testbed}
\vspace*{-0.25 cm}
\end{figure*}


In this paper, we provide an overview of our real-time \SI{28}{\GHz} massive \ac{MIMO} testbed, which includes a hybrid beamforming architecture based on beam selection, as illustrated in \fig\ref{fig:Sys_testbed}. Our testbed constitutes a flexible platform that supports up to 64-antenna/16-\ac{TRx} \ac{BS}, simultaneously serving a maximum of 12 \acp{UE} using \ac{OFDM} in \ac{TDD} mode.



\section{Testbed Architecture}
\label{sec:test_arc}
In this section, we overview the architecture of our testbed. As illustrated in \fig\ref{fig:Sys_testbed}, the proposed testbed is divided into analog and digital subsystems. The physical hardware setup for the digital subsystem is in part identical with the \SI{3.7}{\GHz} 100-antenna/100-\ac{TRx} \ac{LuMaMi} testbed~\cite{mal17}. To extend this to \ac{mmWave}, we have developed the required analog subsystem in-house.
 
The digital subsystem consists of a central control unit and \acp{SDR}~(NI USRP-294xR/295xR), responsible for baseband and \ac{IF} processing. The central control unit has an embedded controller (NI PXIe-8135), which runs \textit{LabVIEW} on a standard Windows 7 64-bit operating system to configure and control the system. LabVIEW provides both host and FPGA programming. To perform \ac{MIMO} processing, \eg, precoding, detection, we use co-processing modules~(FlexRIO 7976R). Also, a reference clock source~(PXIe-6674T) and reference clock distribution network~(Octo-Clock) are included to be able to synchronize the entire \ac{BS}. Each \ac{SDR} contains two \ac{TRx} chains and a Kintex-7 FPGA. The \ac{SDR} basically performs local processing on a per-antenna basis, \eg, \ac{OFDM} processing and reciprocity calibration. Also, it plays a role as an interface to send control signals from the digital to the analog subsystem, where there are two kinds of control signals. One is the signal for \ac{TDD} switching, the other for beam-selection. These control signals are delivered through a 15-pin \ac{GPIO} in each \ac{SDR}.

The analog subsystem includes a 64-element antenna array, a \ac{ClkDist}, \acp{FRECON}, and \acp{FEM} for analog-domain beamforming. For reconfigurability and scalability of the testbed, we designed the \acp{FRECON} and \acp{FEM} in a modular way. Each module has a small number of \ac{TRx} chains, \ie, two per \ac{FRECON} and one per \ac{FEM}. To up/down-convert between \ac{IF} signal from/to the \ac{SDR} and \SI{28}{\GHz} bands, we designed \ac{FRECON} \acp{PCB} that consist of up/down conversion mixers, filters, \acp{DA}, \acp{LNA}, and SPDT switches. One \ac{FRECON} is connected with one \ac{SDR}. The main role of the \ac{FEM} is to switch between four predefined beams, according to the control signal from the digital subsystem. The \ac{FEM}, thus, contains a SP4T switch, and two \acp{FEM} are connected to the \ac{FRECON}. The 64-element antenna array has 16 subarrays. Each subarray, consisting of four antenna elements, plugs in to one \ac{FEM} where one antenna element in the subarray is selected for analog beamforming. In our testbed, the \ac{BS} and \ac{UE}, respectively, has a common \ac{LO} for up/down conversion between \ac{IF} and \SI{28}{\GHz} bands. We employ a \SI{25.5}{\GHz}-\ac{LO}~(PLDRO-25500-10). To amplify and distribute the \ac{LO} signal to multiple \acp{FRECON}, we design the \ac{ClkDist}. \fig\ref{fig:Arch_testbed} illustrates an architecture of 64-antenna/16-\ac{TRx} hybrid beamforming testbed where the quantities of units belonging to each subsystem are also shown.




\begin{figure}[t!]
\centering
\subfigure[ ]
{
\includegraphics[width=3.37in]{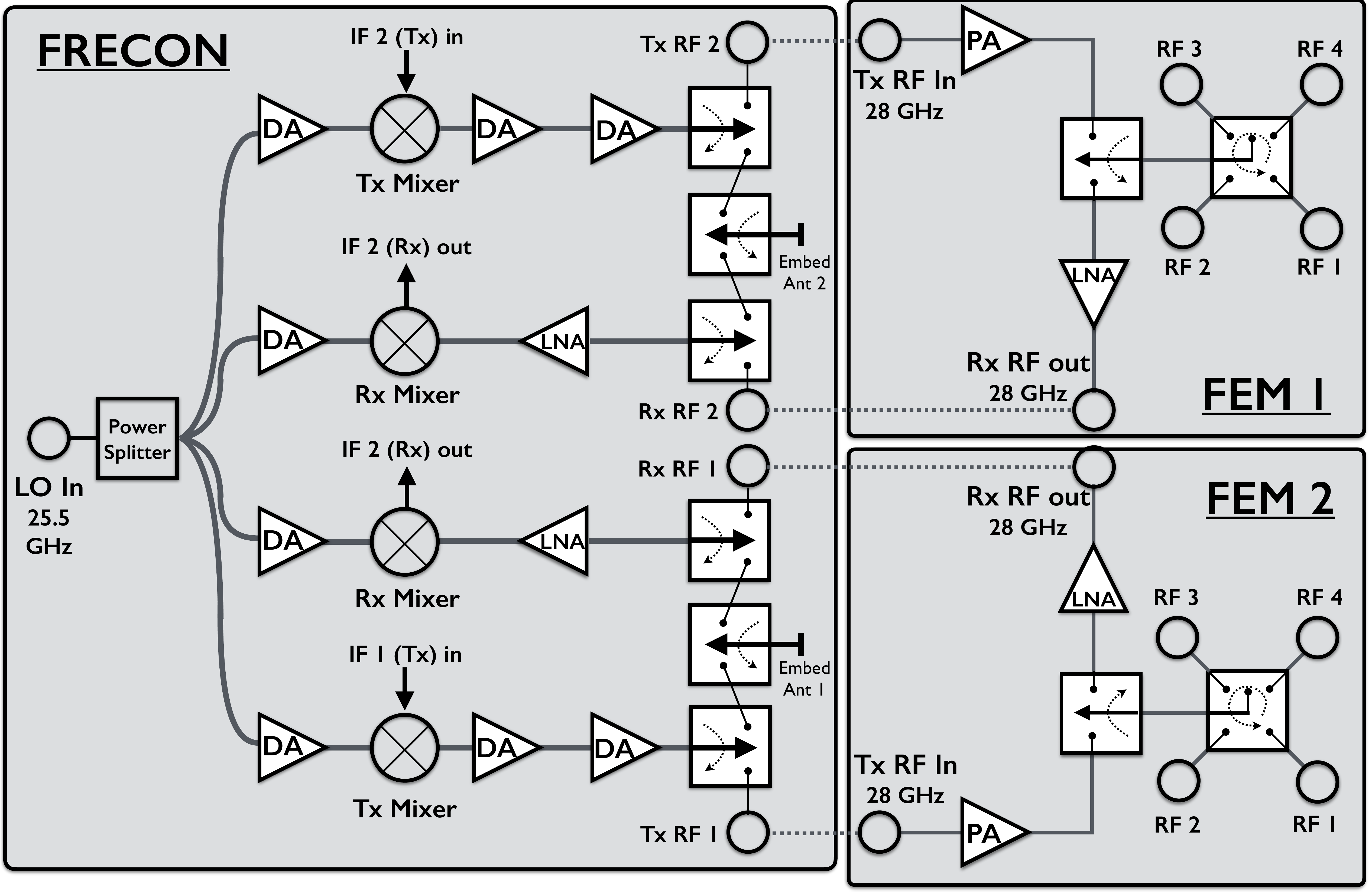}
\label{fig:BD_FRECON_FEM}
}
\subfigure[ ]
{
\includegraphics[width=3.37in]{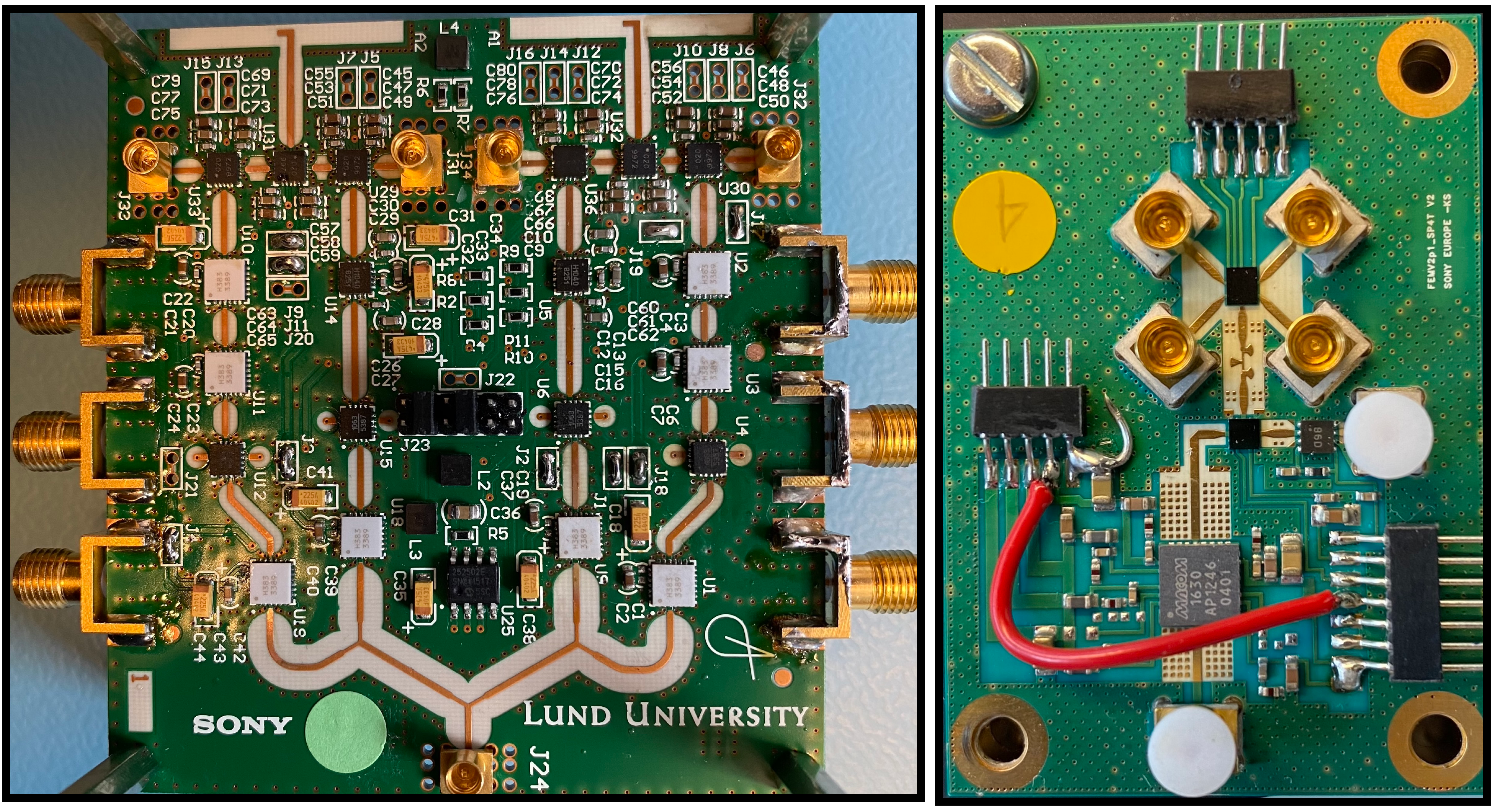}
\label{fig:photo_FRECON_FEM}
}
\caption{FRECON and FEM: (a) block diagram with one FRECON and two FEMs (b) photographs of fabricated FRECON (left) and FEM (right).} 
\label{fig:FRECON_FEM}
\end{figure}
\section{Testbed Design and Implementation}
To perform measurements in a variety of scenarios, sufficient gain of each \ac{TRx} chain is imperative in designing a testbed. Also, for our proposed hybrid beamforming testbed, beam switchability is a key design feature. This section elaborates on the design of our \SI{28}{\GHz} massive \ac{MIMO} testbed and its implementation. 



\subsection{\ac{FRECON} and \ac{FEM}}
MmWave systems are more sensitive to \ac{PA} nonlinearities, compared to conventional systems below-\SI{6}{\GHz}. For the \ac{FRECON} and \ac{FEM}, we focused on an appropriate architecture design and component selection to reduce the \ac{PA} nonlinearities. 

As mentioned in \scn\ref{sec:test_arc}, one \ac{FRECON} has two \ac{TRx} chains, and connects with two \acp{FEM}. A combined block diagram of \ac{FRECON} and \ac{FEM} is shown in \fig\ref{fig:BD_FRECON_FEM}. The \ac{FRECON} contains the same eight \acp{DA}~(HMC383LC4) but has different targets. Four \acp{DA} between an \ac{LO}-input port and mixers is for amplifying the \SI{25.5}{\GHz}-\ac{LO} signal, and the other four \acp{DA} for the \SI{28}{\GHz} \ac{Tx} signal. The mixers~(HMC1063LP3E) are used for up/down-conversion between \ac{IF} and \SI{28}{\GHz} bands, where an \ac{LO} power of more than $10\hspace{0.2em}\rm{dBm}$ is required to operate it. That is the reason why a \ac{DA} for amplifying the \ac{LO} signal is needed for each mixer. The conversion gain of the mixer is around \SI{-10}{\dB}. To compensate this power loss and achieve high output power, the \ac{Tx} chain is equipped with two consecutive \acp{DA}. On the other hand, each \ac{Rx} chain contains an \ac{LNA}~(HMC1040LP3CE) to avoid compression. In the front-end of the \ac{FRECON}, there are SPDT switches~(ADRF5020) for \ac{TDD} switching{\footnote{We integrated two commercial antennas for future work, together with four \ac{RF} ports. Thus, there are a total of six SPDT switches to control all the \ac{RF} inputs/outputs of \ac{FRECON}.}}.  

The employment of \ac{FEM} is to support testing of long-range communications, as well as beam-switching. As depicted in \fig\ref{fig:BD_FRECON_FEM}, one \ac{FEM} contains an additional \ac{PA}~(MAAP-011246) and \ac{LNA}, which have a high power gain. Also, a SPDT switch for \ac{TDD} switching, and a SP4T switch for beam-selection are included. The SP4T switch engages with four \ac{RF} ports, and performs switching or selecting by control signals from digital subsystem. For the beam-switching, the isolation between paths in the switches is crucial. The SPDT and SP4T switches, therefore, are designed so that they both have a high isolation{\footnote{The manufactured SPDT and SP4T switches in the \ac{FEM} yielded around \SI{38}{\dB} and \SI{30}{\dB} isolation, respectively.}} in \ac{mmWave} frequencies. All the components, except the PA, were developed in-house.

\begin{figure}[t!]
\centering
\subfigure[ ]
{
\includegraphics[width=3.39in]{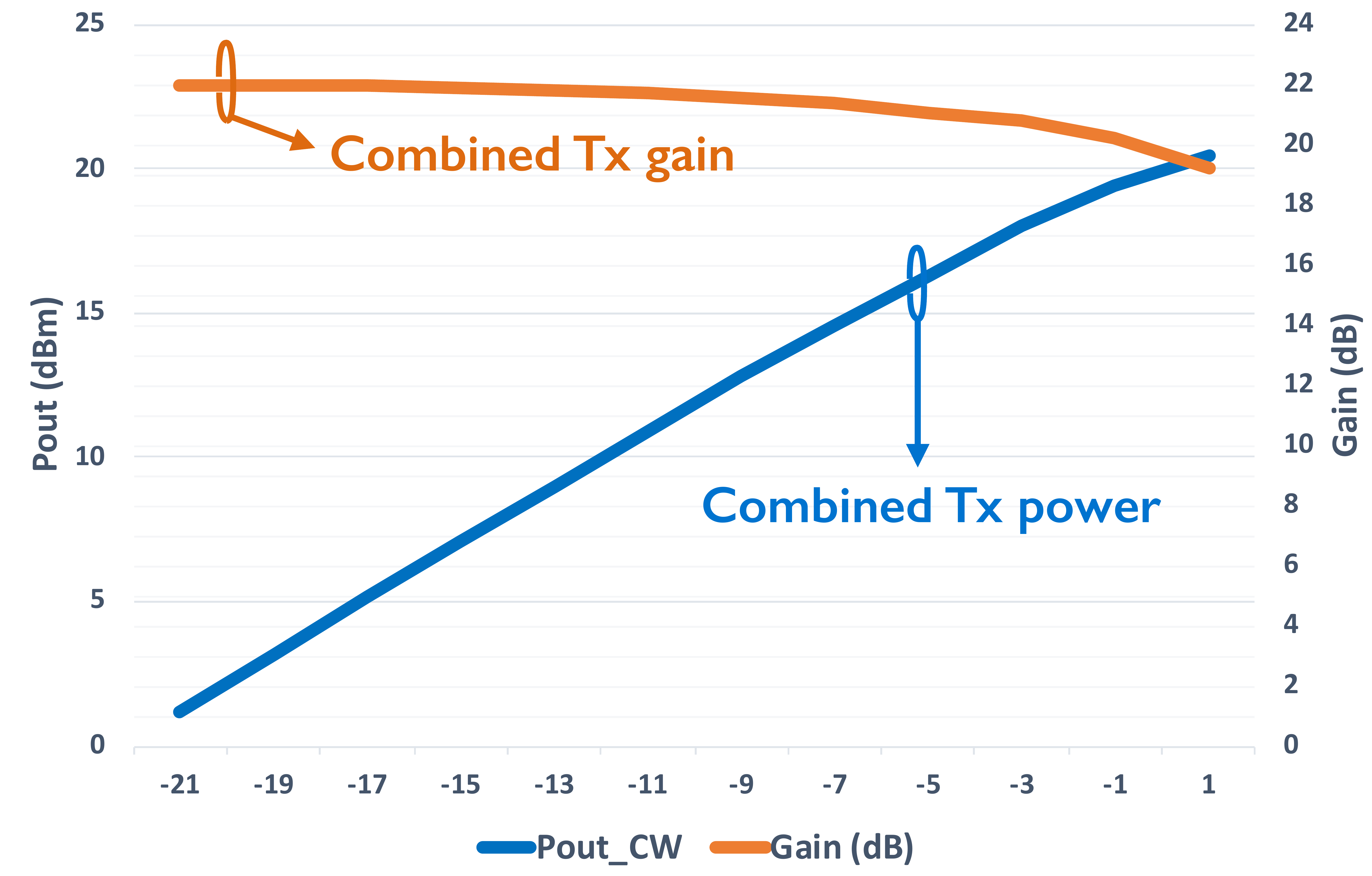}
\label{fig:Tx_gain}
}
\subfigure[ ]
{
\includegraphics[width=2.99in]{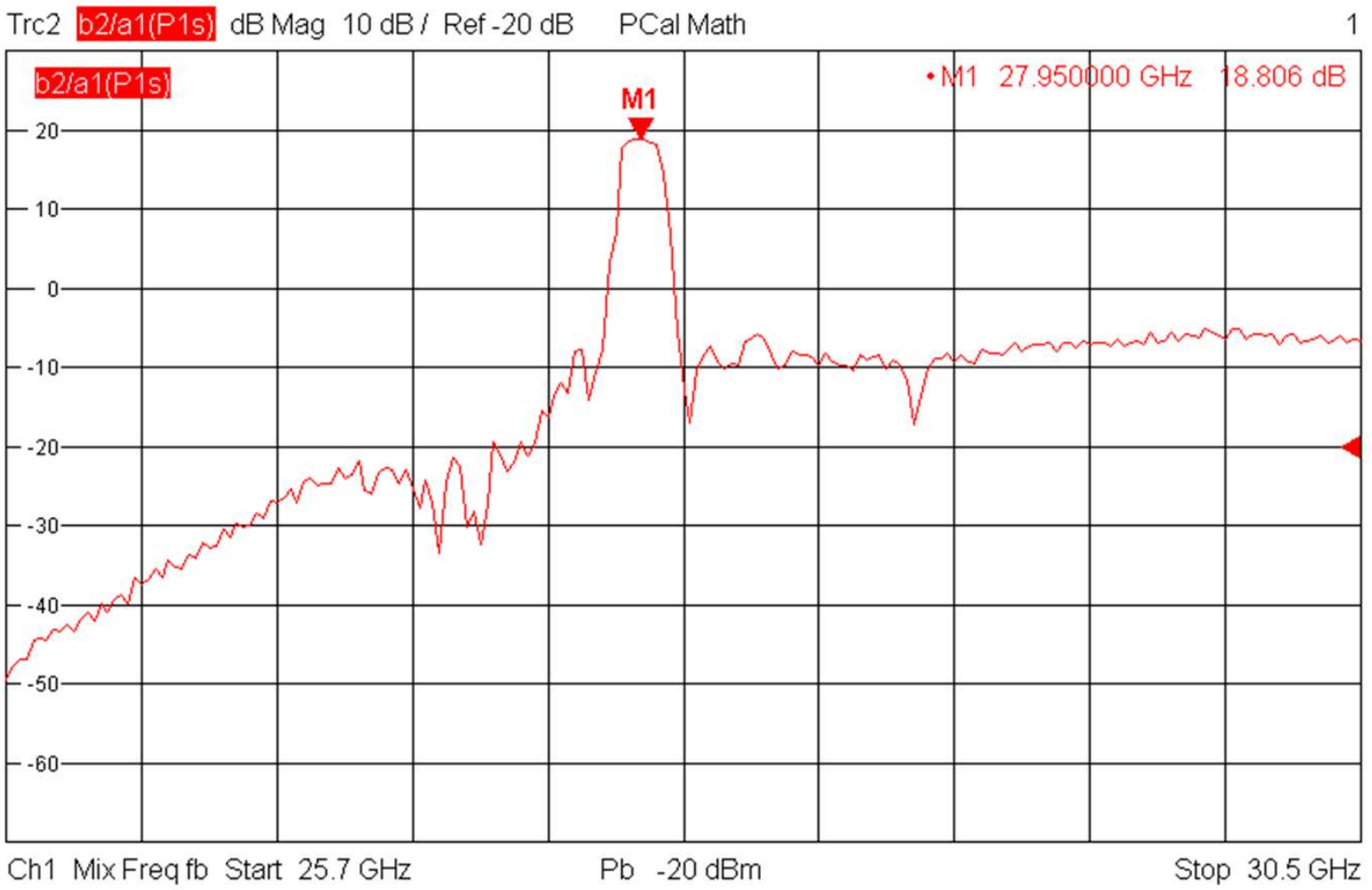}
\label{fig:Rx_gain}
}
\caption{Measurement results of FRECON and FEM: (a) output power level (left vertical-axis) and Tx gain (right vertical-axis) of combined FRECON and FEM (b) Rx gain of combined FRECON and FEM.} 
\label{fig:TRx_gain}
\vspace*{-0.25 cm}
\end{figure}

Photographs of the fabricated \ac{FRECON} and \ac{FEM} are shown in \fig\ref{fig:photo_FRECON_FEM}. Based on the measurements of each module, the \ac{Tx} and \ac{Rx} gains for the \ac{FRECON} are around \SI{9}{\dB} and \SI{7}{\dB}, respectively. For the \ac{FEM}, around \SI{14}{\dB} and \SI{12}{\dB}, respectively, is achieved. \fig\ref{fig:TRx_gain} shows a combined \ac{TRx} gain of \ac{FRECON} and \ac{FEM}. The \ac{Tx} gain in its linear region is around \SI{22}{\dB}, as shown in \fig\ref{fig:Tx_gain}. It delivers a 1-dB gain compression point (P1dB) of $18\hspace{0.2em}\rm{dBm}$. The measured \ac{Rx} gain is shown in \fig\ref{fig:Rx_gain}. Its maximum gain is \SI{18.8}{\dB} at \SI{27.95}{\GHz}. Also, the power consumption of the implemented \ac{FRECON} and \ac{FEM} is \SI{6.3}{\W} and \SI{7}{\W}, respectively.

\subsection{\ac{ClkDist}}
\begin{figure}[t!]
\centering
\includegraphics[width = 3.39in]{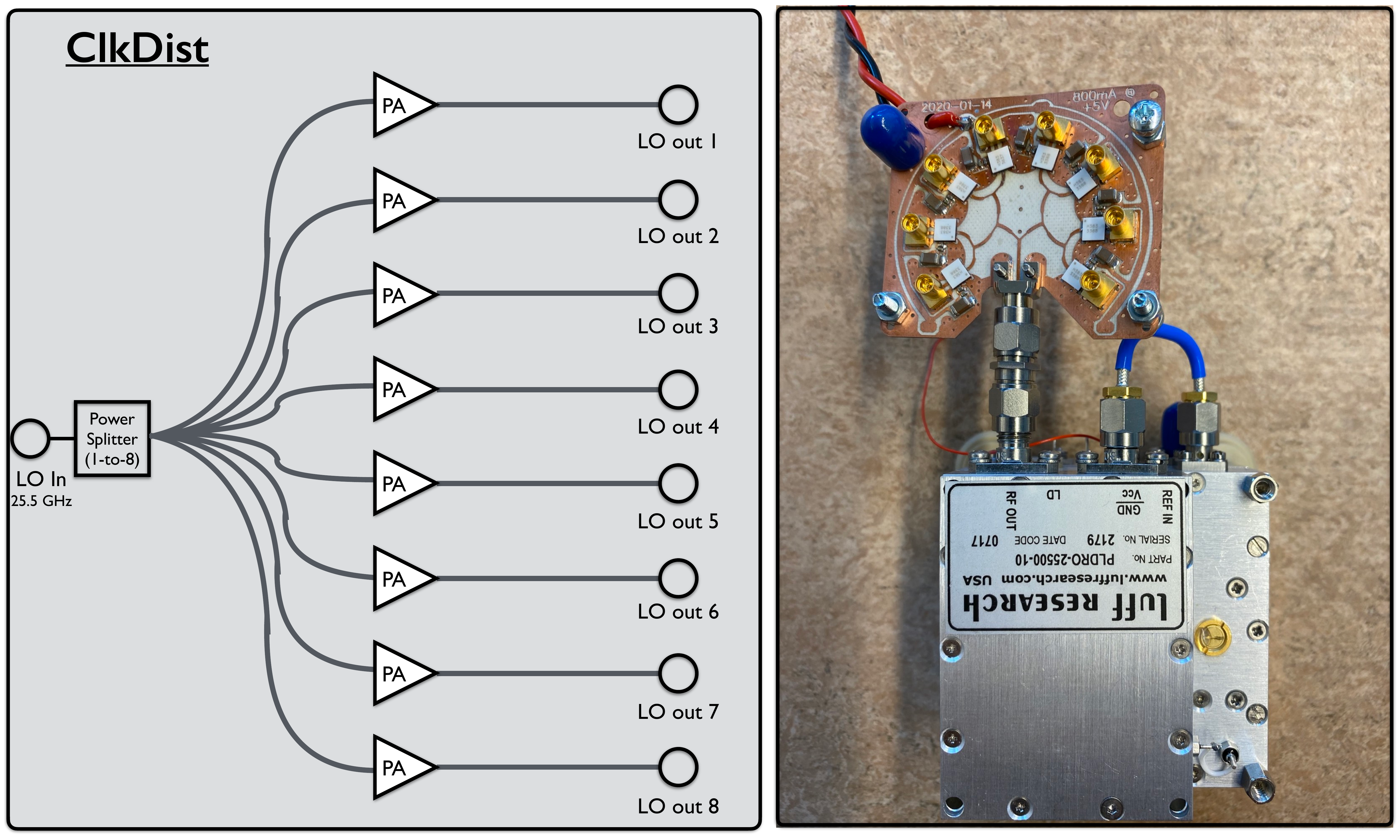}
\caption{Block diagram of ClkDist~(left) and its photograph with a common \SI{25.5}{\GHz}-LO~(right).}
\label{fig:BD_Photo_ClkDist}
\vspace{-0.25 cm}
\end{figure}
The block diagram of the \ac{ClkDist} and its photograph are shown in \fig\ref{fig:BD_Photo_ClkDist}. The \ac{ClkDist} has 1 input and 8 output ports connecting with the \ac{LO} and \acp{FRECON}, respectively. Since the 1-to-8 power splitter causes a power loss of more than \SI{10}{\dB}, each path in the \ac{ClkDist} is equipped with one \ac{DA}~(HMC383LC4)  to meet the input-power requirement of mixers in \ac{FRECON}. The power consumption of the fabricated \ac{ClkDist} and the \ac{LO} is \SI{4}{\W} and \SI{5}{\W}, respectively{\footnote{\acp{LO} operating at \ac{mmWave} frequencies is very sensitive to temperatures. Thus, we adopt a cooling fan to operate our \SI{25.5}{\GHz}-\ac{LO}. Its power consumption is added in the \ac{LO}'s power consumption.}}.

\subsection{Antenna Array}
\begin{figure}[t!]
\centering
\includegraphics[width = 3.39in]{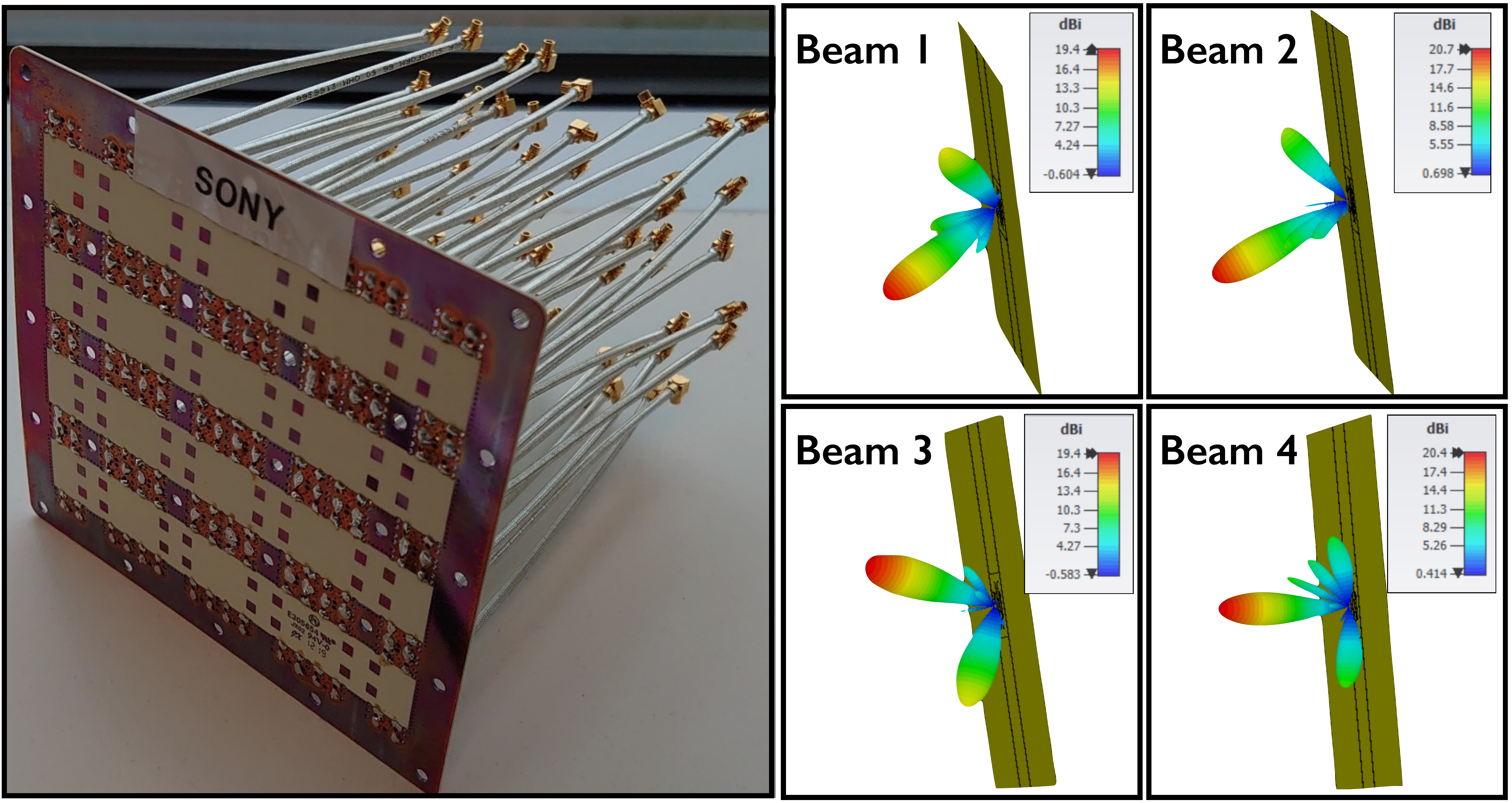}
\caption{Photograph of 64-antenna array~(left) and the beam patterns of 16 subarrays~(right).}
\label{fig:Array_ant}
\vspace{-0.25 cm}
\end{figure}
The planar 64-antenna array, consisting of 16 subarrays, is designed on a three-layer \ac{PCB} using two stacked RO4350B substrates. Each subarray consists of $2 \times 2$ patch antennas with a butler matrix, capable of forming four directional beams. The antenna-element spacing in the subarray is half a wavelength~($\lambda/2$), \ie, \SI{5.5}{\mm}. The spacing between each subarray is $2\lambda$, \ie, \SI{22}{\mm}. The peak gain of single subarray and 16 subarrays, respectively, is $10.1\hspace{0.2em}\rm{dBi}$ and $20.7\hspace{0.2em}\rm{dBi}$. \fig\ref{fig:Array_ant} shows a photograph of the manufactured 64-antenna array and the beam patterns of 16 subarrays.

\subsection{\ac{TDD} and Antenna Switching}
Both \ac{FRECON} and \ac{FEM} have an interface, respectively, to receive DC~(\SI{3.3}{\V}) control signals from the digital subsystem, which are connected with the \ac{GPIO} port. Since the \ac{GPIO} port plugs in to an FPGA embedded in each \ac{SDR}, the DC signal is controllable according to designed blocks in the digital subsystem. For the \ac{TDD} and antenna switching, we implement control units in the digital subsystem, based on the frame structure and baseband functionalities of \ac{LuMaMi} testbed~\cite{mal17}. Since the \ac{LuMaMi} testbed operates in \ac{TDD} mode, its control signal in the digital subsystem can be exploited to deliver to the analog subsystem. Using a regular beam sweeping in the \ac{Rx} mode, channel estimation block computes the channel magnitudes, and returns the antenna index of the highest channel magnitude to the antenna-selection control unit. The system parameters for the developed testbed is summarized in \tbl\ref{table:sys_para}.
\begin{table}[h]
\caption{High-level system parameters}
\renewcommand{\arraystretch}{1.4} 
\centering 
\setlength{\abovecaptionskip}{0pt}
\scalebox{0.9}{
\begin{threeparttable}
\begin{tabular} { >{\centering\arraybackslash}p{4.0cm} 
    >{\centering\arraybackslash}p{4.0cm}}
    \Xhline{3\arrayrulewidth}
    \bf{Parameter}& \bf{Value}\\
\hline
\centering{Carrier frequency} & \SI{27.95}{\GHz}\\ 
\centering{Intermediate frequency} & \SI{2.45}{\GHz}\\
\centering{Sampling frequency} & \SI{30.72}{\MHz}\\
\centering{Signal bandwidth} & \SI{20}{\MHz}\\
\centering{FFT size} & 2048\\
\centering{Antenna-array configuration} & 64 elements\\
\centering{Number of TRx chains} & 16\\ 
\centering{P1dB of each \ac{TRx} chain} & $18\hspace{0.2em}\rm{dBm}$\\ 
\centering{Peak gain of 16 subarrays} & $20.7\hspace{0.2em}\rm{dBi}$\\ 
\Xhline{3\arrayrulewidth} 
\end{tabular}
\end{threeparttable}
}
\vspace*{-.2cm}
\label{table:sys_para}
\end{table}



\section{Initial Results}
This section provides initial results on the link-budget calculation through \ac{OTA} testing. Also, we perform an indoor uplink transmission with 16 \ac{TRx}-chain \ac{BS} and two single-antenna \acp{UE} to validate our testbed design. 

For the link-budget calculation, we used one \ac{FRECON} and one \ac{FEM} for transmission and reception, respectively. To clarify  the \ac{Tx} and \ac{Rx} power of an \ac{IF} signal, a signal generator~(E8257D) is connected to the \ac{FRECON} input of the \ac{Tx} side, and a spectrum analyzer~(FSU50) to the \ac{FRECON} output of the \ac{Rx} side. Based on this setup, we calculate a measured \ac{FSPL} for the distance $d$ between \ac{Tx} and \ac{Rx} antennas, and compare with its theoretical number. \ac{EIRP} is the  hypothetical power radiated by a isotropic \ac{Tx} antenna in the strongest direction and defined as
\begin{align}
\label{eq:eirp}
\mathsf{EIRP}({\rm{dBm}})
= 
P_{\mathsf{tx}}
+
G_{\mathsf{tx}}^{\mathsf{c}}
-
L_{\mathsf{tx}}^{\mathsf{c}}
+
G_{\mathsf{tx}}^{\mathsf{a}}	
\end{align}
where $P_{\mathsf{tx}}$ is the \ac{Tx} power~(\ac{IF} input), $G_{\mathsf{tx}}^{\mathsf{c}}$ the effective \ac{Tx} gain of \ac{FRECON} and \ac{FEM}, $L_{\mathsf{tx}}^{\mathsf{c}}$ the cable loss in the \ac{Tx} side, and $G_{\mathsf{tx}}^{\mathsf{a}}$ the \ac{Tx} antenna gain.
Using the \ac{EIRP}, the measured \ac{FSPL} is
\begin{align}
\label{eq:mPL}
\mathsf{PL}_{\mathsf{m}}({\rm{dB}})
= 
\mathsf{EIRP}
-
(P_{\mathsf{rx}}
-
G_{\mathsf{rx}}^{\mathsf{c}}
+
L_{\mathsf{rx}}^{\mathsf{c}}
-
G_{\mathsf{rx}}^{\mathsf{a}}	)
\end{align}
where $P_{\mathsf{rx}}$ is the \ac{Rx} power~(\ac{IF} output), $G_{\mathsf{rx}}^{\mathsf{c}}$ the effective \ac{Rx} gain of \ac{FRECON} and \ac{FEM}, $L_{\mathsf{rx}}^{\mathsf{c}}$ the cable loss in the \ac{Rx} side, and $G_{\mathsf{rx}}^{\mathsf{a}}$ the \ac{Rx} antenna gain. The theoretical \ac{FSPL} is
\begin{align}
\label{eq:tPL}
\mathsf{PL}_{\mathsf{th}}({\rm{dB}})
= 
20{\rm{log}}_{10}
\Bigg(
\frac{4\hspace{0.06em}\pi\hspace{0.06em}d\hspace{0.06em}f}
{c}
\Bigg)
-
G_{\mathsf{tx}}^{\mathsf{a}}
-
G_{\mathsf{rx}}^{\mathsf{a}}
\end{align}
where $f$ is the carrier frequency, and $c$ is the speed of light. From (\ref{eq:eirp})-(\ref{eq:tPL}), the link-budget calculation results are shown in \tbl\ref{table:link_budget}. It is observed that the measured \acp{FSPL} are quite close to the theoretical ones.
\vspace*{-.2cm}
\begin{table}[h]
\caption{Link-budget calculation results}
\renewcommand{\arraystretch}{1.4} 
\centering 
\setlength{\abovecaptionskip}{0pt}
\scalebox{0.9}{
\begin{threeparttable}
\begin{tabular} { >{\centering\arraybackslash}p{1.0cm} 
    >{\centering\arraybackslash}p{2.0cm}
    >{\centering\arraybackslash}p{2.0cm}
    >{\centering\arraybackslash}p{2.4cm}}
    \Xhline{3\arrayrulewidth}
    \bf{$d$\hspace{0.1em}({\rm{m}})}& \bf{$\mathsf{PL}_{\mathsf{th}}$\hspace{0.1em}({\rm{dB}})}& \bf{$\mathsf{PL}_{\mathsf{m}}$\hspace{0.1em}({\rm{dB}})}& \bf{Gap\hspace{0.1em}($|\mathsf{PL}_{\mathsf{th}}-\mathsf{PL}_{\mathsf{m}}|$)}\\
\hline
\centering{7} & 68.3 & 71.3 & 3\\ 
\centering{8.7} & 70.2 & 72.9 & 2.7\\ 
\Xhline{3\arrayrulewidth} 
\end{tabular}
\end{threeparttable}
}
\vspace*{-.2cm}
\label{table:link_budget}
\end{table}
\begin{figure}[t!]
\centering
\includegraphics[width = 3.39in]{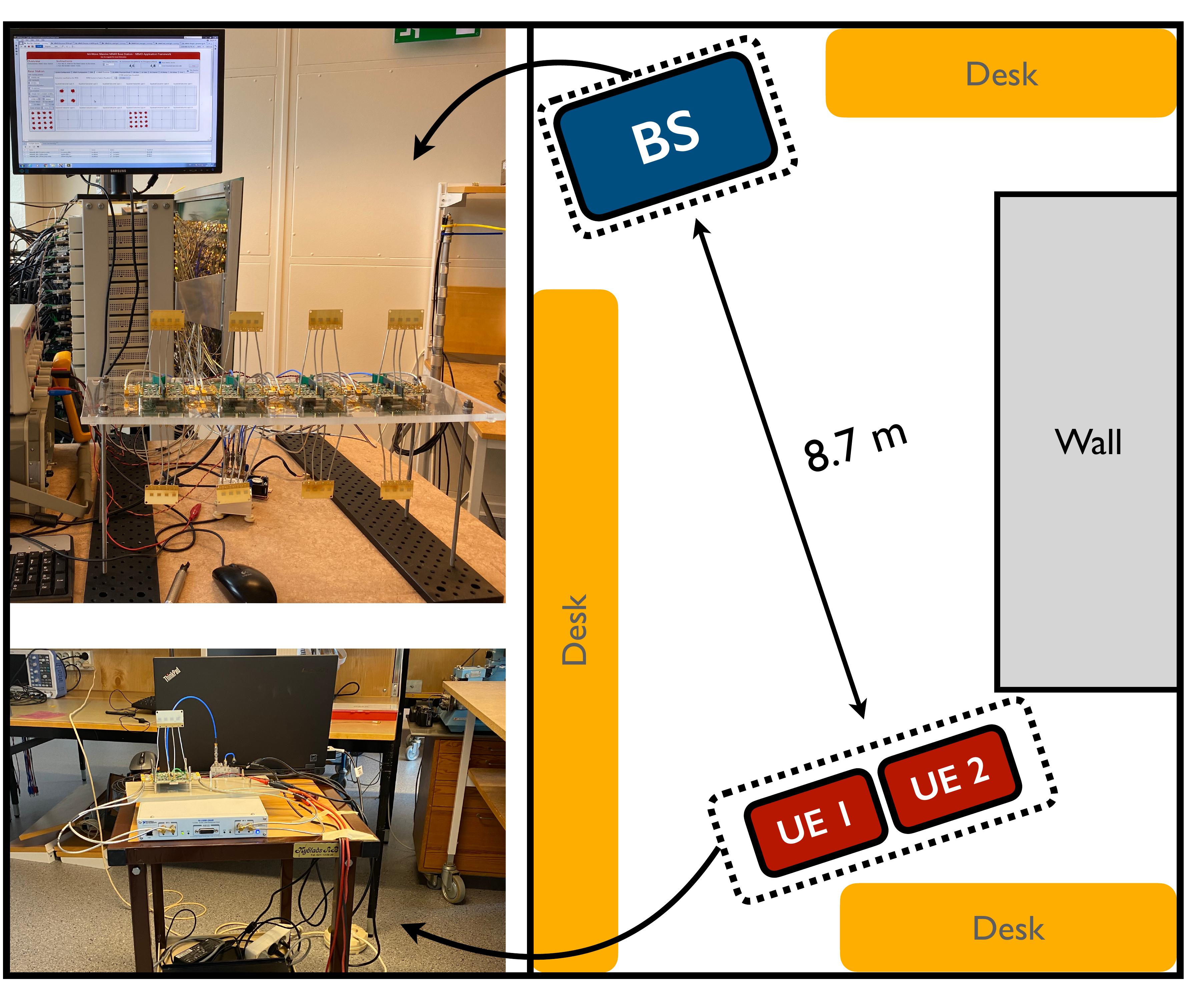}
\caption{Indoor measurement setup in a lab including the positions of the BS and two UEs. There is no obstruction between the BS and the UEs.}
\label{fig:ota_test}
\vspace{-0.25 cm}
\end{figure}
\begin{figure}[t!]
\centering
\includegraphics[width = 3.39in]{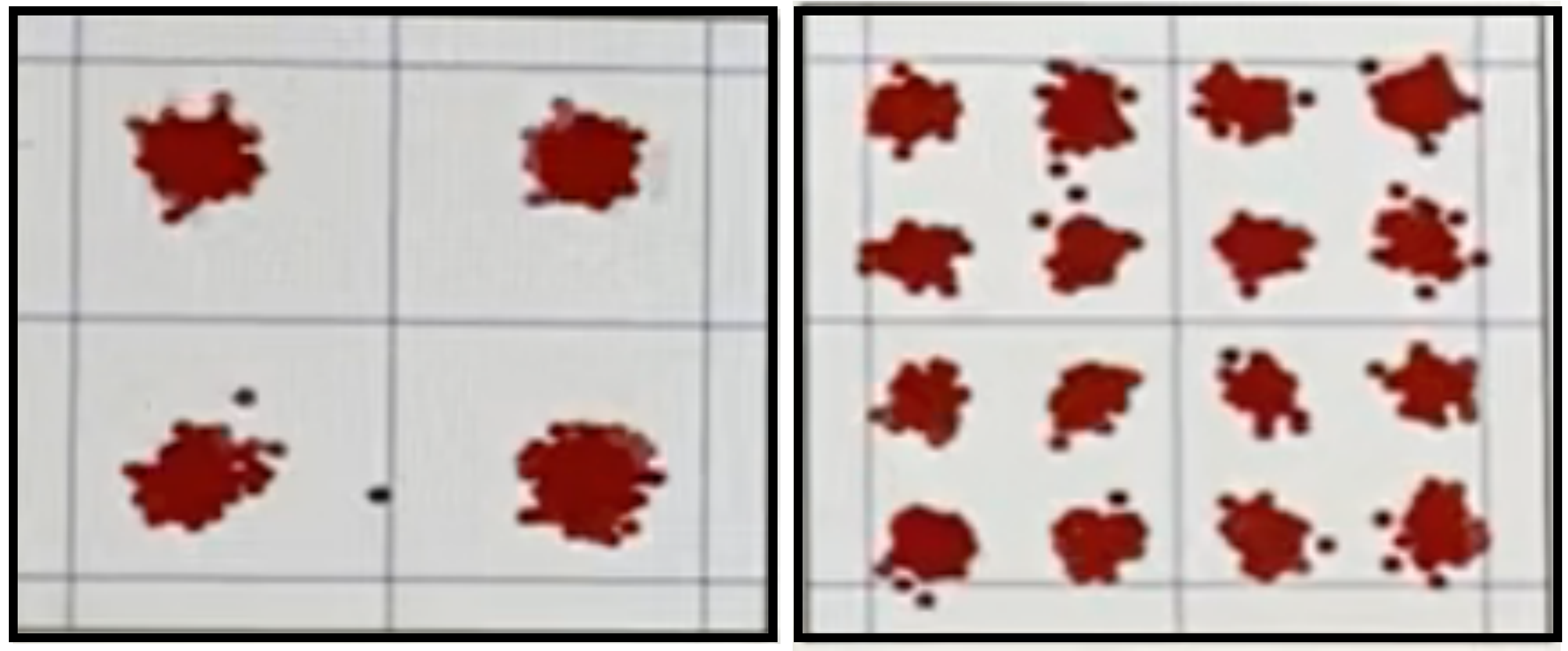}
\caption{Uplink constellations for the indoor experiment when using zero-forcing equalizer at the BS. The UE\hspace{0.1em}1/UE\hspace{0.1em}2 transmit QPSK and 16-QAM, respectively.}
\label{fig:const}
\vspace{-0.25 cm}
\end{figure}

For the indoor test{\footnote{The demo video is available at https://www.youtube.com/watch?v=rgoC6kTJnI8}}, we used a 16-\ac{TRx} fully digital beamforming \ac{BS} and two single-antenna \acp{UE}. The uplink transmission was performed in line-of-sight-like conditions. \fig\ref{fig:ota_test} shows the indoor measurement setup including the positions of the \ac{BS} and \acp{UE}. The distance between the \ac{BS} and the co-located \acp{UE} was \SI{8.7}{\m}. We observed very clear UL constellations. \fig\ref{fig:const} shows captured constellations of the received uplink QPSK~(\ac{UE}\hspace{0.1em}1) and 16-QAM~(\ac{UE}\hspace{0.1em}2), where a zero-forcing equalizer is used at the \ac{BS}.

\section{Conclusion}
Both academia and industry have been making efforts in meeting 5G requirements. To support 5G, massive MIMO and mmWave have each shown strength and potential. Furthermore, it has been known that they are inseparably connected. Realizing mmWave massive MIMO in practice, however, is still an important issue that must be solved. As a viable solution, we have built the real-time \SI{28}{\GHz} massive MIMO testbed with a hybrid beamforming architecture. In this paper, we have provided an overview of our real-time \ac{mmWave}~(\SI{28}{\GHz}) massive \ac{MIMO} testbed, with a hybrid beamforming architecture based on beam-selection, and initial results through \ac{OTA} testing.

\section*{Acknowledgment}
The authors would like to thank Chris Clifton and Kamal K. Samanta at Sony Semiconductor, UK for useful discussions and for providing the \ac{FEM}. In addition, this work is carried out within the Strategic Innovation Program “Smartare Elektroniksystem”, a joint venture of Vinnova, Formas and the Swedish Energy Agency (2018-01534).

\bibliographystyle{IEEEtran}
\bibliography{asilomar_20.bib}  
\input{asilomar_20.acro}

\end{document}